\documentclass[acmtog]{acmart}

\acmSubmissionID{658}

\let\siggraphBbbk\Bbbk
\let\Bbbk\undefined
\let\siggraphmathscr\mathscr
\let\mathscr\undefined

\usepackage[centerfigures,commands,enumerate,environments,noxurl]{AVT}
\usepackage{xurl}
\usepackage{booktabs}
\usepackage{siunitx}
\usepackage{algorithm}
\usepackage{algpseudocode}
\usepackage{docmute}
\usepackage{balance}
\usepackage[capitalize]{cleveref}

\definecolor{hardblue}{RGB}{55,126,184}
\definecolor{hardpurple}{RGB}{152,78,163}

\author{Jingsen Zhu}
\email{jingsen@cs.cornell.edu}
\affiliation{\institution{Cornell University}\country{USA}}

\author{Silvia Sellán}
\email{silviasellan@cs.columbia.edu}
\affiliation{\institution{Columbia University}\country{USA}}

\author{Alexander Terenin}
\email{avt28@cornell.edu}
\affiliation{\institution{Cornell University}\country{USA}}

\let\mathscr\siggraphmathscr
\let\Bbbk\siggraphBbbk

\copyrightyear{2026}
\acmYear{2026}
\setcopyright{cc}
\setcctype{by}
\acmConference[SIGGRAPH Conference Papers '26]{Special Interest Group on Computer Graphics and Interactive Techniques Conference Conference Papers}{July 19--23, 2026}{Los Angeles, CA, USA}
\acmBooktitle{Special Interest Group on Computer Graphics and Interactive Techniques Conference Conference Papers (SIGGRAPH Conference Papers '26), July 19--23, 2026, Los Angeles, CA, USA}
\acmDOI{10.1145/3799902.3811119}
\acmISBN{979-8-4007-2554-8/2026/07}


\begin{document}

\title{A Bayesian Approach for Task-Specific Next-Best-View Selection with Uncertain Geometry}

\begin{abstract}
We develop a framework for task-specific active next-best-view selection in 3D reconstruction from point clouds, by casting the problem in the language of Bayesian decision theory. Our framework works by (a) placing a prior distribution over the space of implicit surfaces, (b) using recently-developed stochastic surface reconstruction methods to calculate the resulting posterior distribution, then (c) using the posterior distribution to carefully reason about which view to scan next. This enables us to perform camera selection in a manner that is directly optimized for the intended use of the reconstructed data---meaning, we reduce uncertainty only in those regions that make a difference in the task at hand, as opposed to prior approaches that reduce it uniformly across space. We evaluate our method across three distinct downstream tasks: semantic classification, segmentation, and PDE-guided physics simulation.
Experimental results demonstrate that our framework achieves superior task performance with fewer views compared to commonly used baselines and prior general uncertainty-reduction techniques. Code for this paper is available at \url{https://github.com/jingsenzhu/BayesianNBV}.
\end{abstract}

\begin{teaserfigure}
\includegraphics[width=\textwidth]{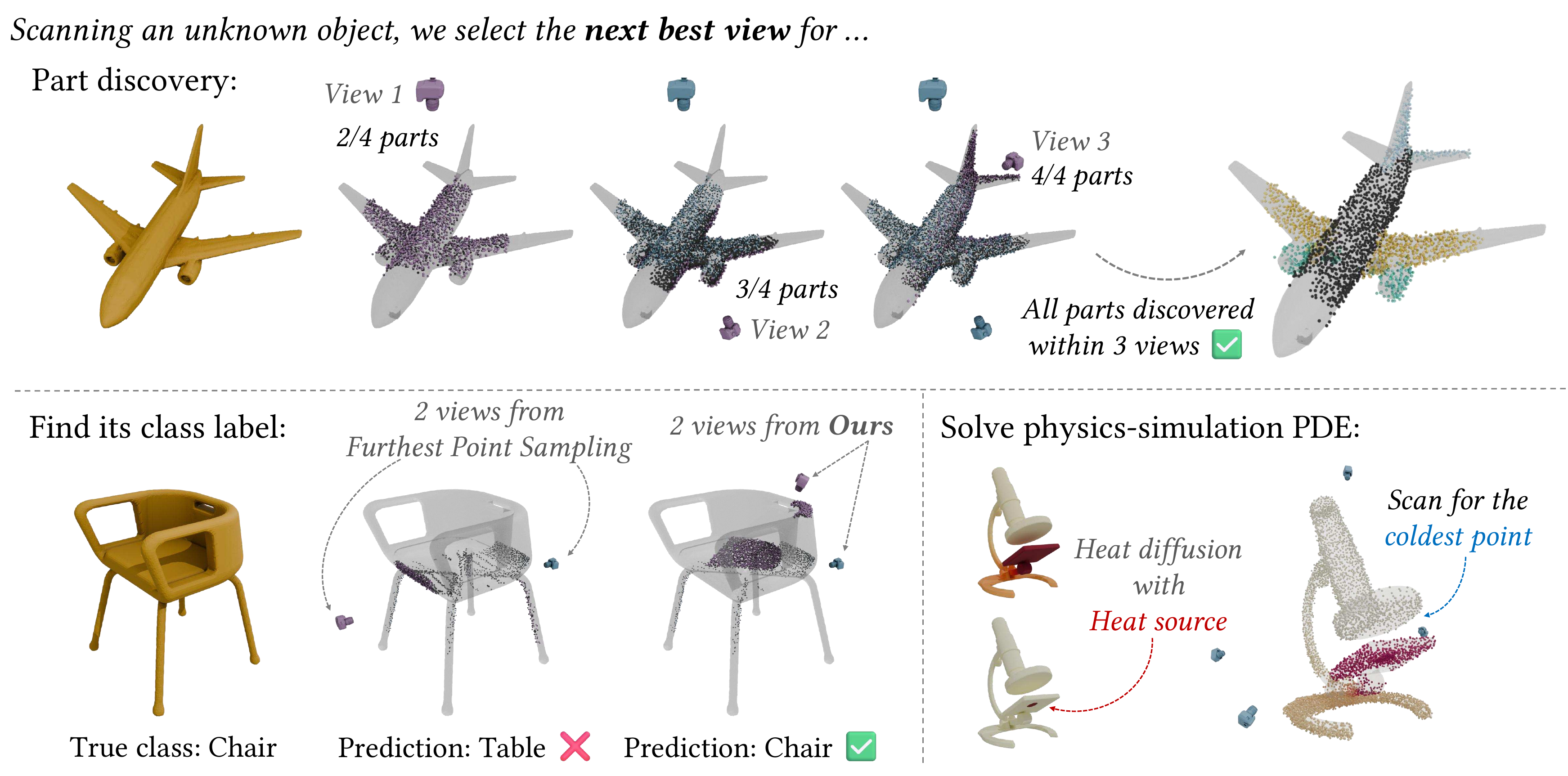}
\caption{We introduce a novel framework for optimizing the next view angle given an incomplete observation of a given object. Our algorithm is \emph{task-specific}, focusing on the regions of the object that matter most for a specific application.}
\label{fig:teaser}
\end{teaserfigure}

\maketitle

\section{Introduction}

\begin{figure*}[t]
\includegraphics[width=\textwidth]{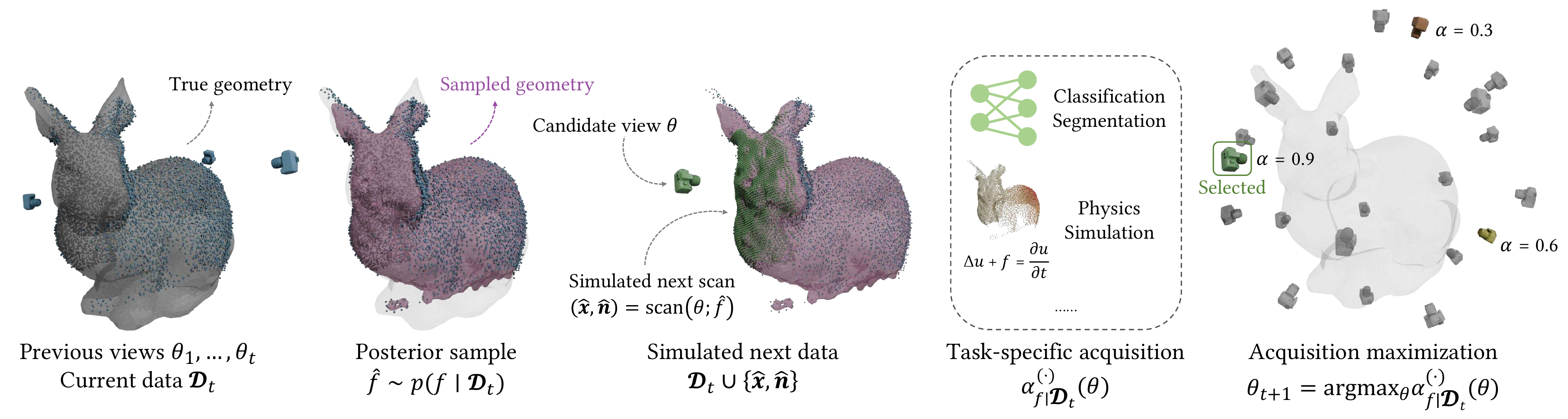}
\caption{Using existing stochastic surface reconstruction techniques, we process an input point cloud (left) and compute a posterior distribution, from which we can estimate the effect a simulated next scan would have on a task-specific acquisition function. We use this to score potential new sensor positions, and choose the highest-scoring one.}
\label{fig:method}
\end{figure*}

Surface reconstruction from 3D point cloud data is a central problem in computer graphics and vision with a wide range of real-world applications, from robotics to medical imaging and geospatial imaging.
In many of these, information about an unknown surface is collected sequentially by one or few sensors, one potentially costly scan at a time.
Often, this information is collected not to produce the most generally accurate reconstruction of the 3D scene---as classical next-best-view approaches attempt to do---but rather to ascertain a specific fact about it, for instance, classify an object, or carry out a specific task, for instance, segment the scene into regions.
Optimizing the next view from which to image the scene to gain the most useful information for our specific goal, which we refer to as \emph{task-specific next-best-view selection}, is thus a critical research question.

Recent work on stochastic surface reconstruction and uncertain geometric representations has put an answer to this question within reach.
Theoretical tools like Gaussian processes, as \citet{sellan2022stochastic} show, can be used to model the uncertainty contained in an incomplete 3D point cloud, are becoming increasingly tractable to work with \cite{holalkere2025stochastic} and have recently found success in neighboring fields \cite{miller2024objects}.

In this paper, we propose exploiting these recent uncertain geometric models to decide the optimal next sensor position in a sequential scanning pipeline oriented towards a specific task.
In particular, we formalize next-best-view selection through the lens of \emph{Bayesian decision theory}, a mathematical framework for decision-making under uncertainty that has already led to state-of-the-art algorithms in other areas, including black-box global optimization \cite{garnett2023bayesian}.
There is a sense in which black-box optimization is similar to surface reconstruction.
Both involve an unknown function: representing the optimization objective, or the unknown surface, respectively.
The algorithm must choose the next location for evaluating the optimization objective, or the next camera viewpoint.
One of our aims, given the strong performance of Bayesian optimization algorithms, will be to thoroughly develop this analogy and turn it into a framework for designing next-best-view selection algorithms.

\begin{figure*}[t]
\includegraphics[width=0.99\textwidth]{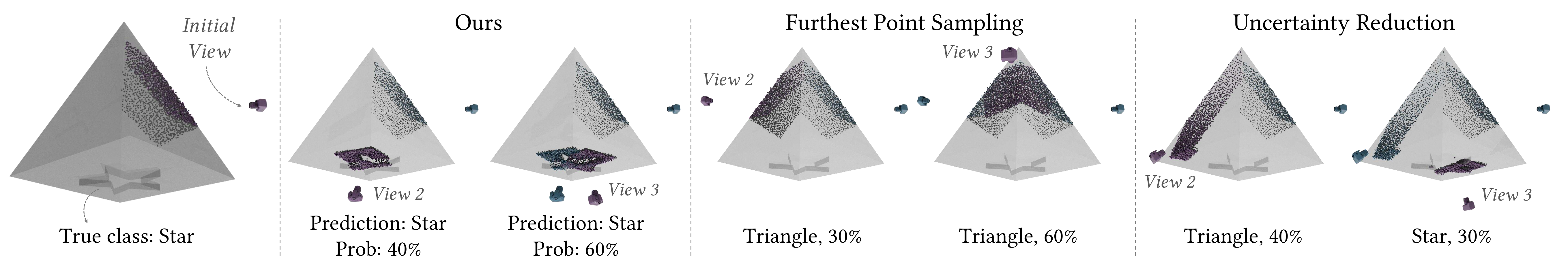}
\includegraphics[width=0.99\textwidth]{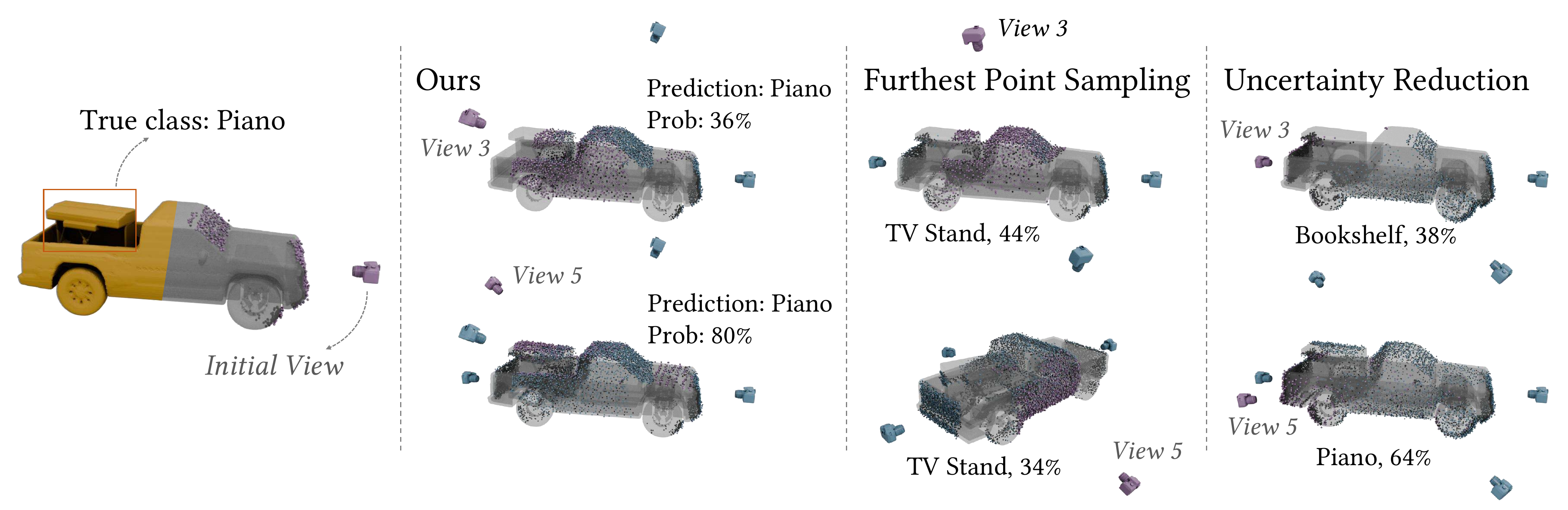}
\caption{Qualitative comparison with FPS and uncertainty reduction in the classification task of the synthetic pyramid and the Truck-ModelNet dataset.}
\label{fig:qual_pyramid}
\label{fig:qual_truck}
\end{figure*}

We propose a first-of-its-kind algorithm that uses Bayesian decision theory to perform \emph{task-specific} next-best-view selection.
This approach builds on recent work in uncertainty quantification for surface reconstruction in order to produce carefully-constructed task-specific optimization objectives, which make it possible to find optimal viewpoints that reduce uncertainty in the locations where it matters most for the specific task (\cref{fig:teaser,fig:method}).
Through prototypical examples, we highlight our algorithm's ability to accept any utility function as a modeling choice, enabling its use in applications from classification (\cref{fig:qual_pyramid}) to partial differential equations (\cref{fig:heat}).

\section{Background and Related Work}

Our work will center on algorithms for the \emph{next-best-view selection} problem.
We will rely on algorithms for \emph{uncertain geometry reconstruction}, and integrate them with techniques from \emph{Bayesian} decision-theory. We now review these three areas.

\subsection{Next-Best-View Selection for 3D Reconstruction}
\label{sec:nbv}

The \emph{next-best-view selection problem} involves choosing a sequence of camera placements in order to learn about some property of an object.
Early approaches in this domain typically operated on voxel-based representations, using geometric heuristics to estimate the information gain of potential views~\cite{isler2016information,bircher2016receding}.
More recently, a number of methods based on neural radiance field representations trained from multi-view RGB images have also been explored \cite{pan2022activenerf,goli2024bayes,jiang2024fisherrf,lyu2024manifold} to achieve convergence with minimal viewpoints.
Scene-level scanning and navigation have also been studied recently, including urban aerial scanning~\cite{tang2025aerial,xiong2025aerial} and indoor scenes~\cite{zhu2025move,chen2025gleam} via structural and semantic understanding.
Our work will focus on next-best-view selection for 3D reconstruction from oriented point cloud observations.
In all of the aforementioned settings, the methods that have been proposed thus far are overwhelmingly task-agnostic, and seek to reduce uncertainty everywhere within the scene.

\subsection{Uncertain Geometry Reconstruction}
\label{sec:spsr}

The most popular method for reconstructing a complete surface from a point cloud is \emph{Poisson surface reconstruction}~\cite{kazhdan2006poisson,kazhdan2013screened}. 
This deterministic approach maps oriented point cloud data into an implicit representation of the surface from which an explicit mesh can be extracted.
While effective for dense inputs, the method becomes ill-posed under partial observations which are consistent with many possible surfaces.

Recent research on \emph{stochastic Poisson surface reconstruction} has addressed this ambiguity by casting reconstruction into the language of \emph{stochastic modeling}, in order to apply techniques from Bayesian learning.
The key idea is to place a prior $p(f)$ over the uncertain implicit function, obtaining a posterior distribution $p(f \given \c{D})$, which describes what was learned about the implicit surface from the point cloud observations $\c{D}$.
Since this formulation is probabilistic, and data is incorporated through conditional distributions, it provides uncertainty quantification across the domain.

For surface reconstruction, \citet{sellan2022stochastic} propose to apply \emph{Gaussian process} models, which are the most widely-used class of Bayesian models for unknown function~\cite{rasmussen2006gaussian}.
In particular, \citet{sellan2022stochastic} introduce models for which the classical Poisson solve is exactly the posterior mean reconstruction, with uncertainty obtained using certain additional linear solves.
Throughout this work, we refer to stochastic models that perform uncertainty quantification for surface reconstruction as models for \emph{uncertain geometry}, to distinguish them from other notions of stochastic geometry that arise in contexts different from ours, such as neural radiance fields~\cite{mildenhall2021nerf}.

A newer variant, called \emph{geometric stochastic Poisson surface reconstruction} \cite{holalkere2025stochastic}, uses ideas from geometric Gaussian processes \cite{borovitskiy2020matern,azangulov2024stationary1,azangulov2024stationary2} to compute $p(f \given\c{D})$ in a single step, without requiring recursive linear system solves.
This variant supports efficient local queries and the generation of posterior samples $\h{f} \~ p(f \given \c{D})$ via pathwise conditioning \cite{wilson2020efficiently,wilson2021pathwise}.
We will adopt it as our primary building block due to its computational tractability: however, our proposed framework is fundamentally model-agnostic, and can integrate with any Bayesian surface reconstruction method that provides a posterior random samples.

\begin{figure*}[t]
\includegraphics[width=0.99\textwidth]{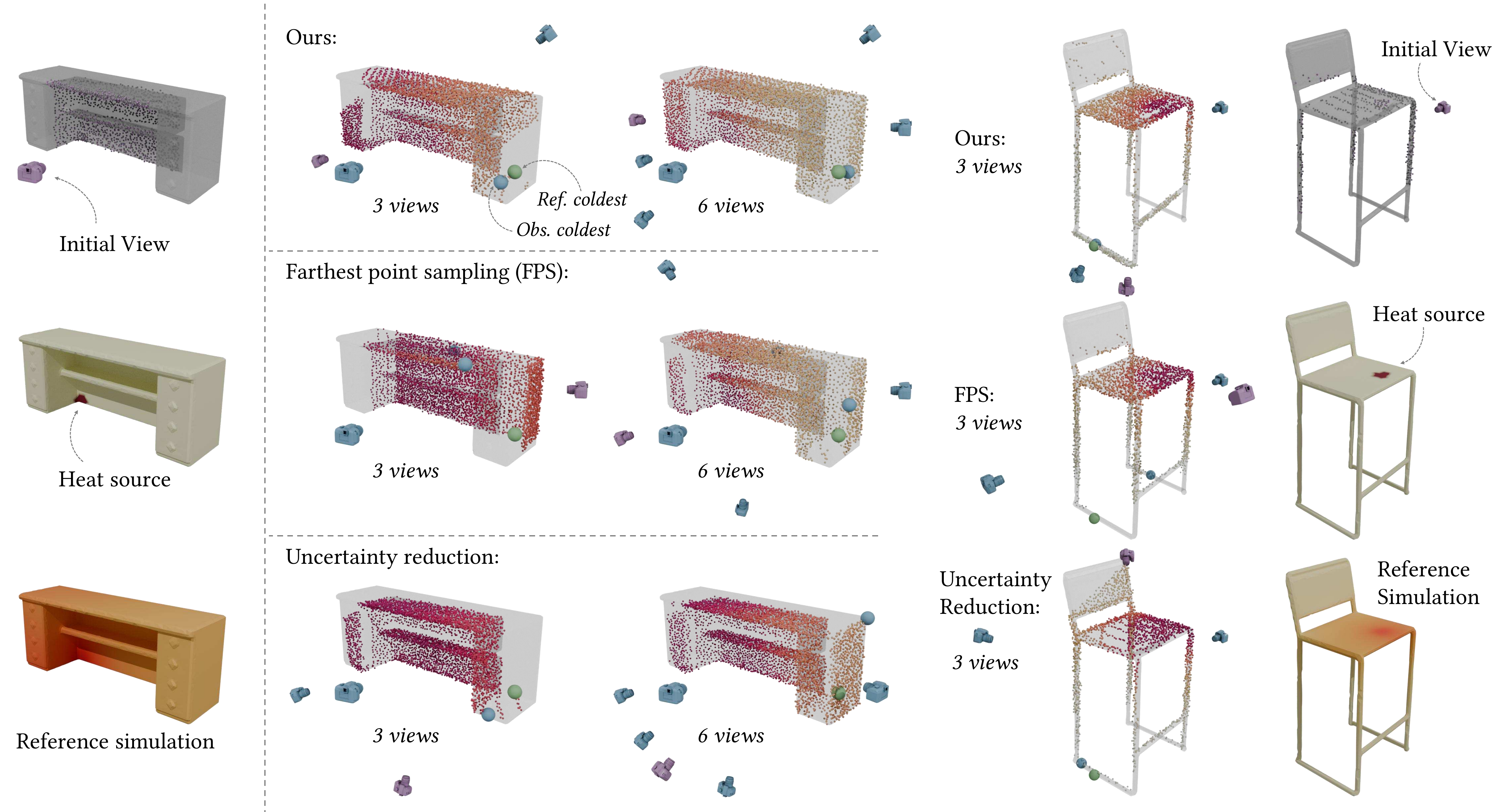}
\caption{Qualitative comparison with FPS and uncertainty reduction in the coldest point discovery task with heat diffusion simulation. The reference coldest point (Ref. coldest) is highlighted in green, and the coldest point simulated from the current observation (Obs. coldest) is highlighted in blue.}
\label{fig:heat}
\end{figure*}

\subsection{Bayesian Decision-making and Bayesian Optimization}
\label{sec:bayesopt}

Through its use of probability, Bayesian learning provides a framework for quantifying uncertainty about a function $f$, through the posterior distribution $p(f \mid \c{D})$.
Given this distribution, \emph{Bayesian decision theory} studies how to effectively use the posterior distribution to further a downstream goal.
One of the most thoroughly-studied Bayesian decision-making setups, called \emph{Bayesian optimization}, involves a goal of finding the global optimum of an unknown function.
This setup may at first seem rather different, but will turn out to be closely-related to the next-best-view selection problem we study, so we review it now.

In black-box optimization, the goal to find the global optimum of an expensive-to-evaluate function $f : X \to\R$, namely 
\[
x^* =  \argmax_{x\in X} f(x)
.
\]
The domain of $f$ is assumed compact, typically $X \subseteq [0,1]^d$.
At each iteration, $t=1,..,T$, the algorithm carefully chooses a point $x_t$, and observes $y_t = f(x_t)$, possibly with noise.
Over time, this allows observations to be collected into a dataset $\c{D}_t = (x_\tau,f(x_\tau))_{\tau=1}^t$.

A Bayesian optimization algorithm works by placing a prior distribution over the objective function.
Given $\c{D}_t$, the algorithm chooses $x_{t+1}$ in two steps.
First, it forms the posterior distribution $f\given\c{D}_t$, by combining the prior and data using Bayes' Rule.
Then, it uses the posterior distribution to form a carefully-constructed function $\alpha_{f\given \c{D}_t} : X\to\R$, called the \emph{acquisition function}, which describes how promising it would be to gather data at a given location.
The algorithm then chooses 
\[
x_{t+1} = \argmax_{x\in X} \alpha_{f\given \c{D}_t}(x)
.
\]
One of the most popular acquisition functions in widespread use is called \emph{expected improvement}, which is defined as
\[
\alpha_{f\given\c{D}_t}(x) &= \bb{E} \max\del{0,(f\given\c{D}_t)(x) - y^*_t}
\\
&= \del{\mu_t(x)\!-\!y^*_t}\Phi\del{\frac{\mu_t(x)\!-\!y^*_t}{\sigma_t(x)}} + \sigma_t(x)\phi\del{\frac{\mu_t(x)\!-\!y^*_t}{\sigma_t(x)}}
.
\]
where $\Phi$ and $\phi$ are the Gaussian CDF and PDF, respectively, $\mu_t$ and $\sigma_t$ are the posterior mean and posterior standard deviation of $f\given\c{D}_t$, also respectively, and $y^*_t = \max_{\tau=1,..,t} y_\tau$.
This acquisition function quantifies how much each location improves on the best point seen so far, averaged over uncertainty quantified by the current posterior distribution.
The second expression helps see how it resolves the explore-exploit tradeoff caused by incomplete information: it does so by prioritizing two kinds of points, namely those whose posterior mean is high, representing good performance, and those whose posterior variance is high, representing uncertainty.

Expected improvement can be derived from first principles.
Suppose that the true objective function is random, and follows the same distribution as the prior used to construct the model.
Then, by standard results from stochastic control, there exists an optimal policy for selecting the next data point, given by a certain intractable dynamic program.
Suppose, at a given time point $t$, we make a \emph{one-step greedy approximation}, and pretend that the decision process will end at time $t+1$, instead of time $T \geq t+1$.
Under this approximation, one can show that the resulting optimal actions maximize expected improvement \cite[Ch. 7]{garnett2023bayesian}.

While our focus will be on improvement-based methods because they generalize well to our setting, we conclude by noting that there are many alternative approaches available, including those based on upper confidence bounds \cite{srinivas2009gaussian}, information-theoretic quantities \cite{hernandez2014predictive,wang2017max}, and Gittins indices \cite{xie2024cost,scully2025gittins}.
We refer to \citet{garnett2023bayesian} for a comprehensive review of Bayesian optimization, and proceed to our formulation.

\section{Next-Best-View as a Bayesian Decision Problem}
\label{sec:method}

We now study next-best-view selection, where our strategy will be to leverage the probabilistic uncertainty quantification capabilities provided by stochastic Poisson surface reconstruction of \cref{sec:spsr} in order to frame scanning as a Bayesian decision problem.
To achieve this, our aim will be to generalize expected improvement of \cref{sec:bayesopt} from the setting of black-box global optimization into the setting of active point cloud scanning.

To do so, we will first define a general Bayesian decision-making framework for camera selection in \cref{sec:framework}, specializing general theory on \emph{expected utility improvement} to the scanning setting.
This will naturally result in \emph{task-specific} approaches, which will only seek to reduce uncertainty in those locations where it makes a difference---in contrast with more traditional approaches, which instead maximize spatial coverage or global uncertainty reduction.
In \cref{sec:tasks}, we demonstrate the flexibility of our approach by deriving task-specific acquisition functions for three distinct scenarios: 3D point cloud classification, semantic segmentation, and physics simulation.

\subsection{Active Scanning as a Bayesian Decision Problem}
\label{sec:framework}

We now formalize active point cloud scanning as a Bayesian decision problem.
Let $f : [0,1]^3 \to [-1,1]$ be an implicit surface representation of the unknown scene.
We assume that information about $f$ is acquired through a \emph{scanning operation} $\f{scan}(\.;f) : SE(3) \to \bb{D}$, where $\theta\in\f{SE}(3)$ represents camera angles, and $\bb{D}$ is a space representing oriented point clouds. 
More precisely, $(\v{x},\v{n}) \in \bb{D}$ consists of a dataset of surface locations $\v{x}$, and surface normals $\v{n}$, of matched but otherwise potentially variable length.
We assume the unknown geometry $f$ is only accessible in a black-box manner through these discrete observations, to mimic 3D scanning applications in which no prior knowledge is assumed about the object.

In a sequential scanning setting, at each time step $t$, we have a set of previously selected viewpoints $\theta_\tau$, $\tau=1,..,t$ and the resulting cumulative dataset $\c{D}_t = \U_{\tau=1}^t \f{scan}(\theta_\tau;f)$, where the union symbol denotes concatenation.
Our objective is to carefully select a subsequent viewpoint $\theta_{t+1}$.

To do so, like in Bayesian optimization, we will place a prior distribution $p(f)$ over the implicit surface, obtaining a posterior $p(f \given \c{D}_t)$.
We extend the scanning operation to also allow \emph{simulated} scans using the posterior distribution, denoted by $\f{scan}(\.;f\given\c{D}_t)$.

We will work with several acquisition functions, a number of which will require an additional ingredient: a \emph{utility function} $u : \bb{D} \to \R$.
This function describes how useful the information contained in a partial scan is for the task at hand.
We will therefore propose a number of utility functions suitable for various tasks, but defer this to \Cref{sec:tasks}.

Given a utility function, we propose to select the next view according to the \emph{expected utility improvement} acquisition function $\alpha^{(u)}_{f\given\c{D}_t} : SE(3) \to \R$, which scores the potential value of a candidate camera $\theta$, and is defined as
\[ \label{eq:eui}
\alpha^{(u)}_{f\given\c{D}_t}(\theta) = \E \max\del[1]{0, u(\c{D}_t \u \f{scan}(\theta; f\given\c{D}_t))- u(\c{D}_t)}
.
\]
This acquisition functions is an example of the general class of improvement-based acquisition functions: see \citet[Ch. 5 and Ch. 6]{garnett2023bayesian} for an introduction to this class.
Using it, we select the next view by solving the optimization problem 
\[
\theta_{t+1} = \argmax_{\theta \in \Theta} \alpha^{(u)}_{f\given\c{D}_t}(\theta)
.
\]
This procedure can intuitively be understood as follows: since $f$ is unknown, we cannot compute the actual utility of a future scan.
Instead, we use the uncertainty from the posterior distribution to simulate potential observations on random surfaces that represent what might happen, and calculate the expected benefit of the scan across these samples.

\paragraph{Practical Implementation}

In practice, we use Monte Carlo sampling to approximate the expectation which defines expected utility improvement, as---unlike the expected improvement acquisition function of \cref{sec:bayesopt}---it cannot generally be calculated in closed form.
For this, we require a stochastic model $p(f\given\c{D}_t)$ which provides the ability to generate random function samples $f\given\c{D}_t$ that can be evaluated at arbitrary locations.
Geometric stochastic Poisson surface reconstruction natively supports this capability using pathwise conditioning, up to a minor approximation: see \citet{holalkere2025stochastic,wilson2020efficiently,wilson2021pathwise} for details.

Given the ability to evaluate $\alpha^{(u)}_{f\given\c{D}_t}$, the next challenge is how to optimize it.
We consider two strategies for maximizing $\alpha^{(u)}_{f\given\c{D}_t}$ over the camera parameter space: \emph{discrete candidate search} and \emph{multi-start gradient-based optimization}.

Discrete candidate search is arguably the simplest possible approach, and solves the optimization problem approximately along a finite candidate pool, which is selected randomly using Monte Carlo or quasi Monte Carlo sampling.
The candidate pool is generated in an easy and task-agnostic manner compared to the actual scanning operation, e.g., uniformly sampled from a bounding sphere.
Though it may appear to be too-simple, this approach has successfully been used previously in other works on next-best-view selection~\cite{holalkere2025stochastic,pan2022activenerf,jiang2024fisherrf}.

Multi-start gradient-based optimization, in contrast, works with the full search space.
However, it generally requires $\alpha^{(u)}_{f\given\c{D}_t}$ to be implemented in an automatic differentiation framework.
In Bayesian optimization, acquisition function optimization landscapes are typically highly non-convex and may have multiple local optima---but, nonetheless, often amenable to gradient-based methods in practice.
The standard approach is to initialize optimization at multiple random locations: see \citet[Appendix A.3]{lin2023sampling} for example implementation details.
Following this, with this approach, we select the final viewpoint $\theta_{t+1}$ of maximum acquisition value across all converged trajectories.

\cref{alg:plan} provides pseudo-code for our Bayesian next-best-view selection algorithm, for the case that discrete random search is used for maximizing the acquisition function.

\begin{algorithm}[t!]
\caption{Bayesian next-best-view selection \hfill\scriptsize(discrete candidate search)}
\label{alg:plan}
\begin{algorithmic}
    \Require Number of cameras to scan $N$, an initial camera $\theta_1$
    \Require Number of candidates $N_\theta$, number of samples $S$
    \Ensure Camera views $\vartheta = \{\theta_i\}_{i=1}^N$, scanned point cloud $\c{D}$
    \State $\vartheta\gets\{\theta_1\}$
    \State $\c{D}\gets\f{scan}(\theta_1;f)$ \Comment{The initial scan}
    \For{$i := 2,\dots,N$}
        \State $\Theta\gets$ \Call{GenerateCandidates}{$N_\theta$}
        \State $\h{f}_1,..,\h{f}_S \~ p(f\mid\c{D})$ \Comment{Draw samples from the posterior}
        \State $\theta_i\gets\displaystyle\argmax_{\theta\in\Theta}\frac{1}{S}\sum_{s=1}^S \max\del{0,u\del{\c{D} \u \f{scan}(\theta; \h{f}_s)} - u(\c{D})}$ 
        \State \Comment{Maximize acquisition function via simulated scan}
        \State $\vartheta\gets\vartheta\u\{\theta_i\}$ \Comment{$\theta_i$ is the chosen camera at step $i$}
        \State $\c{D}\gets\c{D}\u\f{scan}(\theta_i;f)$ \Comment{The actual scan at step $i$}
    \EndFor
\end{algorithmic}
\end{algorithm}

\subsection{Task-specific Utility and Acquisition Functions}
\label{sec:tasks}

The general Bayesian framework introduced in \cref{sec:framework} allows for the tailoring of camera selection to the specific requirements of the downstream application, through the selection of utility and acquisition functions.
We now define several such functions, ranging from overall coverage to application-specific objectives.

\subsubsection{3D Point Cloud Classification}
We first consider the task of 3D point cloud classification.
The objective is to identify the correct class label of the underlying scene with the minimum number of viewpoints.
We assume we are given a pre-trained classifier $\Psi$, which maps point clouds $\c{D}$ into probability distribution over $C$ classes.
Our goal is to guide the scanning process to reduce uncertainty of the classifier, with the aim of improving efficiency by avoiding irrelevant regions that do not make a difference for predicted classes.
We consider two strategies for this.

\paragraph{Expected Entropy Reduction}
The simplest strategy is to choose the utility function to represent uncertainty of the classifier's prediction, which can be quantified by the (negated) Shannon entropy of the softmax output.
Letting $\v{p} = \Psi(\c{D})$, this is
\[
u_E(\c{D}) = \sum_{c=1}^C p_c \log p_c,
.
\]
which defines its respective expected utility improvement acquisition function $\alpha^{(\f{E})}_{f\given\c{D}_t}$.

\paragraph{Expected Cross-Entropy}
As a complementary strategy, we propose an alternative acquisition function which seeks points that cause predictions to shift directly, without an intermediate specification of a utility function.
This is based on the expected cross-entropy between the current and future predicted distributions.
Let $\v{p}^{(t)} = \Psi(\c{D}_t)$ and $\v{p}^{(t+1)} = \Psi(\c{D}_t \u \f{scan}(\theta; f\given \c{D}_t))$, and define
\[
\alpha^{(\f{CE})}_{f\given\c{D}_t}(\theta) = -\E\sum_{c=1}^C p^{(t)}_c \log p^{(t+1)}_c
.
\]
Maximizing this acquisition function directly encourages the selection of views that significantly alter the classifier's current prediction.
In early stages with limited data, this may prevent the classifier from getting stuck in an over-confident state, by forcing it to explore views that contradict its predictions.
As the classification converges to the true label, we expect the cross-entropy to decrease, as additional views yield no new information for classification.

\subsubsection{3D Semantic Segmentation and Part Discovery}

Next, we consider the task of semantic segmentation, where the objective is to discover the various semantic parts of an object with minimal observations.
This task necessitates a semantic exploration strategy, as information may often by non-uniformly distributed across a scene.

We define the problem as follows: given a pre-trained segmentation network $\Phi$, which processes a point cloud $\c{D}$ and outputs a probability distribution $\v{p}_i$ over $C$ semantic classes for each point $i$.
For each point $i$, the predicted part label is $y_i = \argmax_{c=1,..,C} p_{ic}$.
In practice, predictions can be noisy: thus, we define a semantic part $c$ as being \emph{discovered} only when the number of points assigned to that class, $N_c = \sum_{i=1}^N \1_{y_i = c}$, exceeds a user-defined threshold $N_{\f{target}}$.
This threshold allows the user to specify the desired level of granularity or part-completeness for the task.

To guide the scanning process, we define a utility function $u_{\f{S}}(\c{D})$ based on the network's predictive state.
We propose a \emph{soft-count} utility function that utilizes the full softmax distribution to account for model confidence, given by
\[
u_{\f{S}}(\c{D}) = \sum_{c=1}^C \tanh \del{\frac{1}{N_{\f{target}}} \sum_{i=1}^{N} p_{ic}}
\]
from which we can again define the respective expected utility improvement acquisition function $\alpha^{(\f{S})}_{f\given\c{D}_t}$.

The term $\sum_{i} p_{ic}$ represents a differentiable \emph{soft count} of the points belonging to part $c$, preserving the uncertainty information encoded in the network's output.
The $\tanh$ activation function is critical here as it introduces a \emph{saturation} effect: once a part is sufficiently discovered---meaning, the count exceeds $N_{\f{target}}$---the utility gain for that class plateaus.
This effectively shifts the acquisition focus toward underrepresented or undiscovered semantic regions, prioritizing viewpoints that contribute novel semantic information over those that merely add redundant points to already identified parts.

\subsubsection{Physics-Informed Scanning: Heat Diffusion}

Moving beyond semantic and geometric objectives, we consider a task where view planning is driven by the physical properties of the reconstructed scene.
Physics simulations frequently require solving partial differential equations involving the Laplace--Beltrami operator: as a representative example, we consider the heat equation
\[
\pd{u}{t} = \Delta u + f
\]
where $u(x,t)$ represents the temperature at point $x$ and time $t$, and $f(x, t)$ represents a heat source, which is considered time-invariant $f(x)$ in our case.
Given an initial temperature $u_0 = u(x,0)$, a discrete Laplacian matrix $\m{L}$, and a time step $h$, the evolution of the temperature field can be discretized using an implicit Euler scheme
\[
(\m{I} - h\m{L})\v{u}_{k+1} = \v{u}_k + h\v{f}
\]
where $\v{u}$ and $\v{f}$ are vectors defined over the point set.
While $\m{L}$ is typically computed via the cotangent weight formula for meshes~\cite{pinkall1993computing}, point cloud Laplacian methods~\cite{sharp2020laplacian,pang2024neural} support direct evaluation of the operator on discrete unstructured point clouds. 

We define a specific downstream objective: given a known heat source and initial temperature distribution, identify the coldest point on the object after time $T$ using the minimum number of camera views.
For a scanned point cloud $\c{D}$, let $\v{u}^{(T)}$ be the vector obtained by iteratively solving the heat equation for $T/h$ steps.
Using this, define a utility function $u_{\f{H}}(\c{D})$ as the negation of the minimum simulated temperature
\[
u_{\f{H}}(\c{D}) = -\min(\v{u}^{(T)})
.
\]
This gives rise to an expected utility improvement acquisition function $\alpha^{(\f{H})}_{f\given\c{D}_t}$, which seeks the next viewpoint that maximizes the expected decrease in the minimum observed temperature.
Intuitively, this acquisition function prioritizes scanning regions that act as heat sinks, or are thermally isolated from the heat source.

\subsubsection{Geometric Exploration via Surface Coverage}

We finally demonstrate that the traditional task-agnostic 3D reconstruction objective, which typically aims for scene coverage and global scene uncertainty reduction~\cite{sellan2022stochastic,pan2022activenerf}, is also compatible within our framework via an acquisition function based on a unidirectional Chamfer distance between the potential next scan and the existing point cloud to quantify the novelty of a candidate viewpoint, namely
\[
\alpha^{(\f{CD})}_{f \given \c{D}_t}(\theta) = \E  \sum_{\v{x} \in \f{scan}(\theta; f \given \c{D}_t)} \min_{\v{y} \in \c{D}_t} \norm{\v{x} - \v{y}}_2^2
.
\]
Maximizing this objective yields viewpoints whose expected observations are spatially distant from the current dataset $\c{D}_t$, thereby maximizing the discovery of new geometric features.

\begin{figure}[t!]
\includegraphics[trim={0 0 3 0},clip]{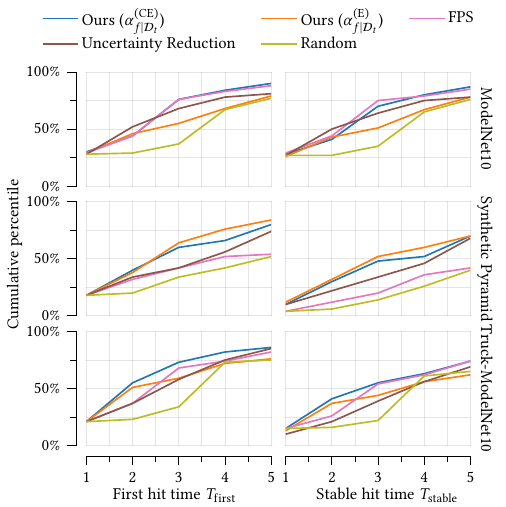}
\caption{Quantitative comparison on the classification of the ModelNet10 (first row), Synthetic Pyramid (second row), and Truck-ModelNet10 (third row) datasets. We use discrete candidate search as the optimization strategy in all comparisons here with the same candidate pool as the baselines.}
\label{fig:merge-class}
\end{figure}

\subsection{Implementation Details}
\label{sec:implement}

Our framework is implemented in PyTorch, with optimization performed using automatic differentiation.
We provide details on the implementation of scanning, stochastic Poisson surface reconstruction, and other aspects in the supplementary material.

For viewpoint optimization, we implement two strategies.
The \emph{discrete candidate search} utilizes a pool of $120$ cameras uniformly distributed on a bounding sphere via a Fibonacci lattice~\cite{gonzalez2010measurement}.
For \emph{multi-start gradient descent}, we initialize six starting cameras randomly sampled on the bounding sphere, optimizing the acquisition function using the Adam optimizer~\cite{kingma2014adam} with learning rate $\eta = 0.01$ for $20$ iterations.
We compare two camera parameterizations: (1) a constrained 2-DOF model restricted to the bounding sphere looking at the origin, and (2) a full 6-DOF pose using a 6D continuous rotation representation~\cite{zhou2019continuity} and translation vector.

The task-specific models and packages we use are as follows.
We utilize PointNet++~\cite{qi2017pointnet++} for classification and segmentation. 
For the heat diffusion simulation, we adopt the approach proposed by \citet[Sec. 5.7]{sharp2020laplacian}, which applies a tufted cover Laplacian operator to a $K$-nearest-neighbor local triangulation of the point cloud as an estimate for the point cloud's Laplacian.

\section{Experiments}

We evaluate our framework across all downstream tasks defined in \cref{sec:tasks}. 
Quantitative and qualitative results demonstrate that our task-specific approach consistently outperforms established baselines, validating the framework's efficiency in tailoring observations to specific applications. 
We also evaluate our framework on the traditional, non-task-specific 3D reconstruction task in the supplementary material.
Additional experimental details are documented in the supplementary material.
We now present these results.

\begin{figure}[t!]
\includegraphics[width=\linewidth]{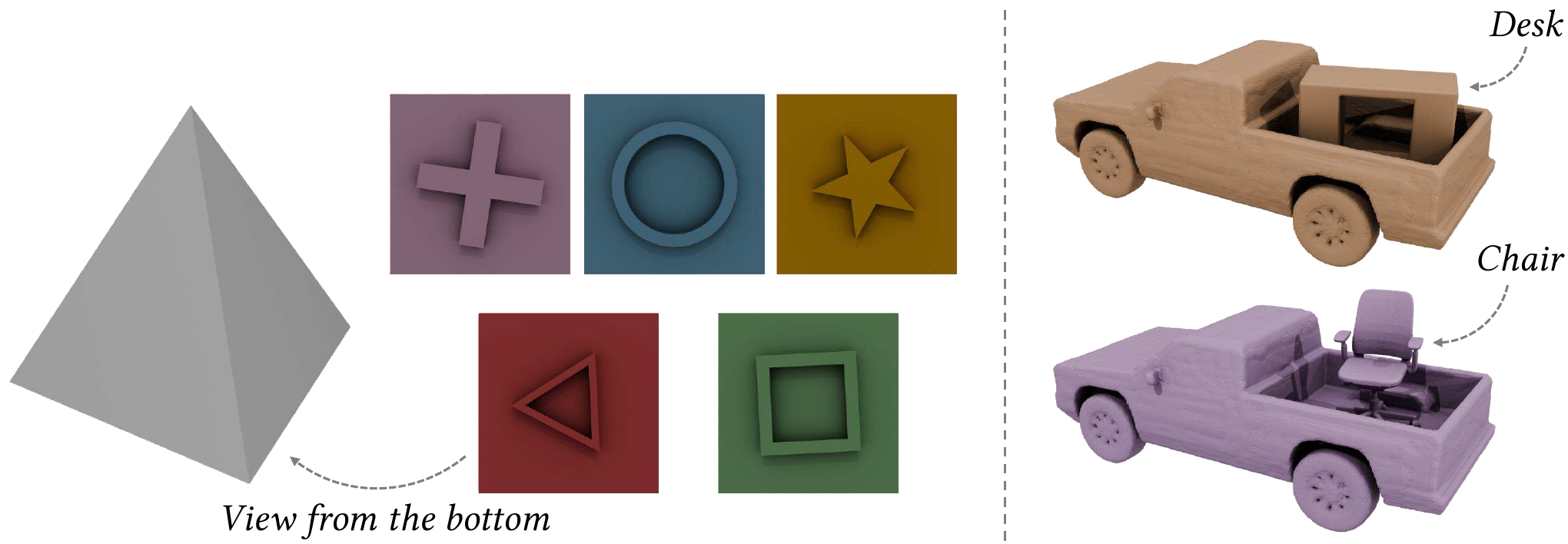}
\caption{Demonstration of our synthetic pyramid and Truck-ModelNet10 datasets. The synthetic pyramid dataset consists of 5 classes: \emph{cross}, \emph{ring}, \emph{star}, \emph{triangle}, and \emph{square}.}
\label{fig:pyramid_demo}
\end{figure}

\paragraph{Baselines}
We compare against three strategies: (1) \emph{furthest point sampling (FPS)}~\cite{eldar1997farthest}, which selects viewpoints at the maximum distance from previous cameras to ensure spatial coverage, (2) \emph{uncertainty reduction}~\cite{holalkere2025stochastic}, which uses the same stochastic model we do, but picks viewpoints of maximum geometric uncertainty along the principal ray, and (3) \emph{random search}, which selects points at uniformly at random from the candidate pool. 
To ensure fair comparisons, all baselines and our discrete search strategy utilize the same candidate pool. While FPS and UR target task-agnostic geometric exploration, our method optimizes for the specific utility of the downstream task, so the key questions will center around how task-specificity affects performance.

\subsection{3D Classification}
\label{sec:result_class}

In this experiment, we evaluate on 10 randomly selected scenes per class, resulting in $10\x N_{\t{classes}}$ scenes per dataset. 
To ensure a fair comparison, all methods utilize the same initial camera and discrete candidate pool. We evaluate two metrics: (1) \emph{first hit time}, denoted $T_{\t{first}}$, which is the earliest step with a correct prediction, and (2) \emph{stable hit time}, denoted $T_{\t{stable}}$, which is the earliest step after which the correct prediction remains consistent for all remaining steps.

\paragraph{Initial sanity checks: classification in settings where global features suffice}
We first evaluate on ModelNet10~\cite{wu20153d}, plotting the \emph{cumulative percentile} of scenes achieving first correct and stable hits within 1--5 steps, shown in \cref{fig:merge-class} on the first row.
Higher curves indicate faster convergence. 
On this benchmark, our method performs comparably to baselines. 
Examining the output, we see that distinguishing ModelNet10 classes, such as \emph{chair} vs. \emph{desk}, relies primarily on \emph{global structure}, which is well-captured by coverage-based heuristics like FPS. 
This indicates task-agnostic exploration is sufficient for coarse classification, motivating evaluation on more complex scenes where \emph{local discriminative features} are critical.

\begin{figure}[t!]
\includegraphics{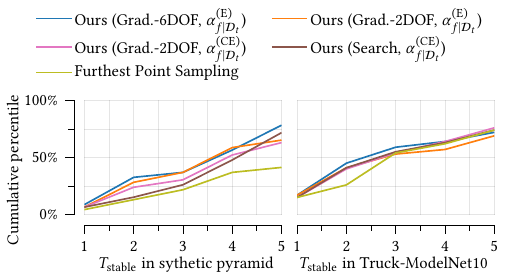}
\caption{Comparison of the stable hit time $T_{\f{stable}}$ in the synthetic pyramid and Truck-ModelNet10 datasets, including between our discrete candidate search and multi-start gradient-based optimization strategies.}
\label{fig:grad}
\end{figure}

\paragraph{Classification in settings where local features are needed}
To evaluate the framework's ability to identify local discriminative features, we created two synthetic datasets where global structure is insufficient for classification:
(1) \emph{Synthetic Pyramid}, which features five distinct patterns, shown in \cref{fig:pyramid_demo}, imprinted on a pyramid's base. 
Accurate classification requires viewpoints oriented toward the bottom, as side views of the lateral faces are non-informative. 
(2) \emph{Truck-ModelNet10}: which places ModelNet10 objects on the cargo bed of a ShapeNet truck~\cite{chang2015shapenet}.
Both the truck and its cargo are randomly oriented. 
To classify the cargo, the scanning agent must specifically target the truck's bed rather than its exterior.

Quantitative results, given in \cref{fig:merge-class}, second and third rows, show that our method significantly outperforms baselines, particularly in early stages, for instance $t=2$.
While baselines struggle with task-agnostic exploration, our cross-entropy acquisition consistently identifies informative regions---namely, the pyramid base and truck cargo area---yielding faster convergence.
Qualitative results, shown in \cref{fig:qual_pyramid,fig:qual_truck}, further confirm that our framework selectively scans these high-utility surfaces, focusing on the most-important areas.

\paragraph{Acquisition comparisons} 
For these tasks, we observe that the entropy-based acquisition function can underperform relative to cross-entropy.
We observe that this stems from a common explore-exploit failure: if the initial scan yields a high-confidence but incorrect prediction---for instance, misidentifying a sofa cushion as a bed---then future viewpoints may change the prediction, but will not reduce uncertainty further, as it is already small.
This may cause the method to prematurely stop exploring.
In contrast, cross-entropy explicitly encourages shifts in prediction, forcing the system to seek views that challenge the current guess, leading to more robustness and faster convergence to the true label.
For a utility-based approaches, this example therefore reveals a key limitation: it is possible for the utility function to interact poorly with upstream model errors, and it must be chosen carefully to ensure performance.

\begin{figure*}[h!]
\includegraphics[width=0.95\textwidth]{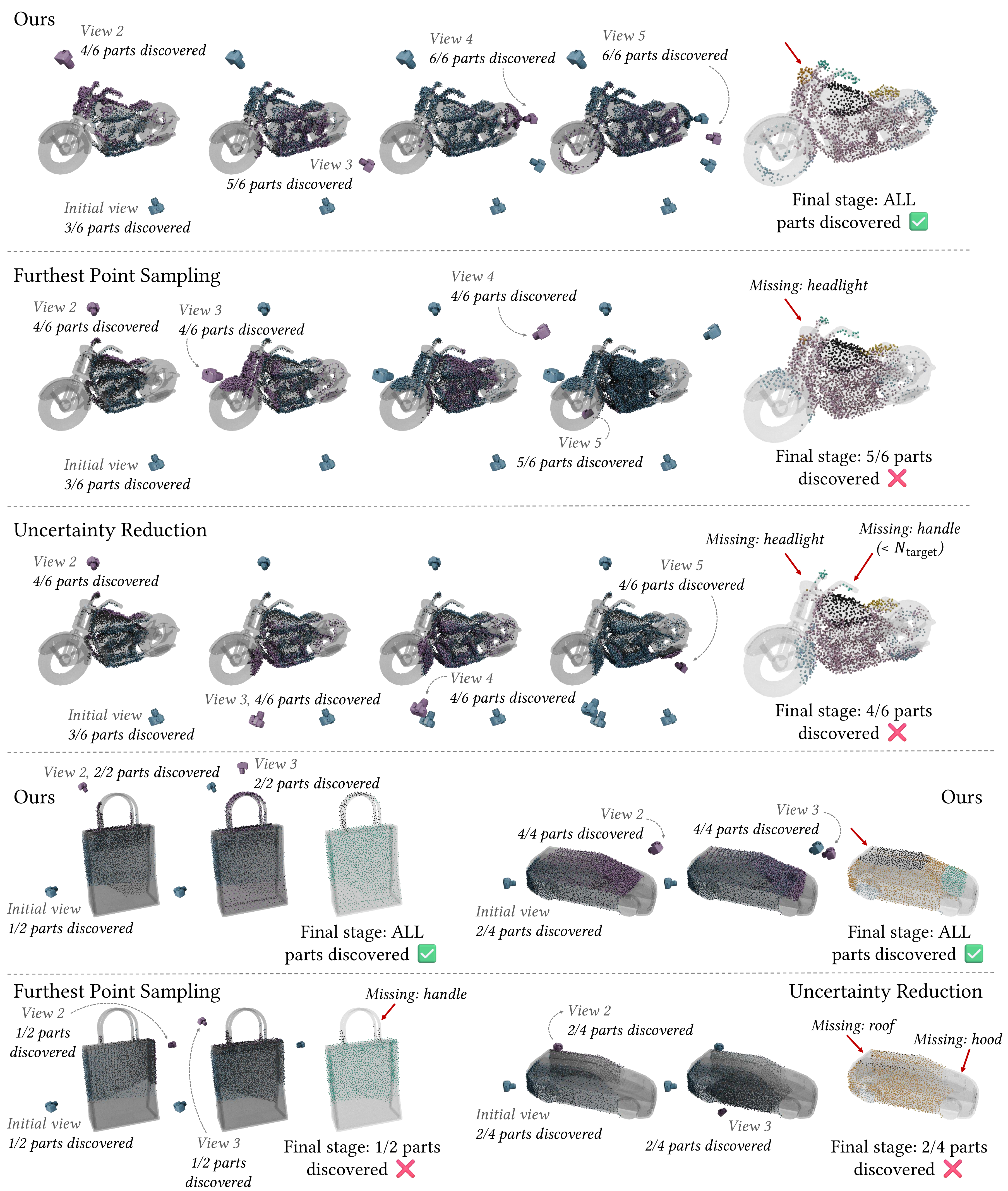}
\caption{Qualitative comparison with FPS and uncertainty reduction in the segmentation task of ShapeNet dataset. The viewpoint selected by the current step and the corresponding scanned points are visualized as \textcolor{hardpurple}{purple}, while the previous viewpoints and existing point cloud are visualized as \textcolor{hardblue}{blue}. The final part segmentation result is shown in the left-most column for each scene.}
\label{fig:qual_partseg}
\end{figure*}

\paragraph{Discrete search vs. gradient-based optimization}
We compare discrete candidate search against multi-start gradient-based optimization, shown in \cref{fig:grad}, for our two benchmark datasets. 
We exclude cross-entropy from the 6-DOF case, because this acquisition functions should be used with constraints, otherwise optimization will divergence due to preferring views that point completely away from the scene.
We see that gradient-based approaches yield only slightly better results than discrete search on the Pyramid dataset and comparable performance on Truck-ModelNet10.
This behavior is different from typical experiences in Bayesian optimization \cite{garnett2023bayesian}, where much-higher-dimensional problems are typical.
This may occur, in part, because many scanning tasks are not sensitive to small variations in camera angles, enabling discrete search to obtain reasonable coverage.
We therefore recommend this strategy as the default choice for most applications.

\begin{table}[t!]
\caption{The average number of cameras required for discovering all parts of the object over 80 test shapes, under different $N_{\f{target}}$ criteria. The best result is \textbf{bold}, second-best is \underline{underlined}, and third-best is \textit{italic}. Less is better.}
\label{tab:segmentation}
\begin{tabular}{lccccc}
\toprule
$N_{\f{target}}$ & 20 & 40 & 60 & 80 & 100 \\\midrule
Ours (Search) & 2.93 & \textit{3.16} & \underline{3.43} & \underline{3.67} & \textbf{3.75} \\
Ours (Gradient-2DOF) & \underline{2.81} & \textbf{3.04} & \textbf{3.28} & \textbf{3.59} & \underline{3.87} \\
Ours (Gradient-6DOF) & \textbf{2.70} & \underline{3.11} & \underline{3.43} & \textit{3.69} & \textit{3.95} \\
FPS & \textit{2.91} & 3.30 & 3.65 & 3.85 & 4.11 \\
Uncertainty & 3.25 & 3.31 & 3.54 & 3.71 & 4.13 \\
Random & 3.29 & 3.42 & 3.65 & 3.93 & 4.10 \\
\bottomrule
\end{tabular}
\end{table}

\subsection{3D Part Segmentation}
\label{sec:result_seg}

For this task, we work with the same scanning configurations as those used in classification. 
We consider a subset of ShapeNet~\cite{chang2015shapenet} comprising 80 scenes across 16 categories. 
To evaluate the efficiency of part discovery, we define $T_{\t{discovery}}$ as the minimum number of viewpoints required to ensure each semantic class contains at least $N_{\t{target}}$ points. 
We report results across five $N_{\t{target}}$ thresholds to evaluate robustness across varying granularity requirements.

\paragraph{Semantic selectivity}
\cref{tab:segmentation} compares our utility-driven methods (discrete and gradient-based) against baselines.
Our framework consistently achieves part discovery with fewer views, with performance gains increasing alongside $N_{\t{target}}$.
This shows that while task-agnostic methods may incidentally capture large semantic regions, our approach specifically targets under-represented parts.

Qualitative results, in \cref{fig:qual_partseg}, highlight this semantic selectivity: our method successfully identifies fine-grained components, such as the motorbike's headlight or the handbag's handle. 
In contrast, geometric baselines often overlook these regions as they contribute little to global surface area, demonstrating the necessity of task-aware planning for detailed semantic understanding.

\subsection{Heat Diffusion Simulation}
\label{sec:result_heat}

For our final task, we evaluate our framework's ability to find the \emph{coldest point} on an object via active scanning using heat diffusion simulation. 
Since the point cloud Laplacian operation given by \cite{sharp2020laplacian} is not (automatically) differentiable, we only consider discrete candidate search for this task. 
\cref{fig:heat} presents qualitative comparisons on objects with complex topologies, visualizing both the simulated temperature distribution on the scanned point cloud and the estimated coldest point relative to ground truth.

Accurate heat diffusion simulation requires a comprehensive understanding of heat transport paths, which are often dictated by hidden or internal structures, such as the underside of a desk or the intricate joints of a chair. 
While geometric baselines prioritize visible surface area, our physics-informed utility function, and resulting acquisition function, proactively identify and scan these critical structural components. 
In \cref{fig:heat}, our method resolves the object's connectivity within a limited number of viewpoints, leading to a significantly more accurate temperature field and a precise estimation of the thermal extrema. 
This demonstrates that our framework can effectively bridge the gap between geometric acquisition and downstream functionality.

\section{Conclusion}

In this work, we introduced a first-of-its-kind task-specific next-best-view selection algorithm for point cloud data, based on combining ideas from stochastic Poisson surface reconstruction with Bayesian decision theory.
On both quantitative and qualitative comparisons, we showed the algorithm to consistently outperform established baselines in settings where the scanning process must find specific local features in order to correctly perform the task.

\subsection{Extensions and Future Work}
In our work, we rely on Gaussian process models, due to both their recent use in stochastic surface reconstruction and popularity in other decision-making settings such as Bayesian optimization.
We believe similar ideas could be applied in other model classes that may enable richer kinds of uncertainty quantification.
Similarly, while we only considered tasks involving point clouds, one could generalize our pipeline to broader kinds of geometry processing.
We see these as promising future work.

Additionally, while we mainly evaluate on object-level scanning, our method provides a generalized framework that can be naturally extended to scene-level settings by carefully designed acquisition functions, which typically require the camera to scan the interior structure of the scene.
Another notable potential direction is to extend our method to non-camera-based scanning settings, e.g., FMRI, to capture the interior structures of an object for various downstream applications. 
We leave these to future work.

\begin{acks}
We thank Ramin Zabih, Steve Marschner and David Bindel for insightful discussions about this project. The first author would like to thank Chuanruo Ning for lending his GPU for some experiments.
The Geometry and the City lab at Columbia University is supported by generous gifts from nTop, Adobe, Dandy and Braid Technologies, and by a Dreamsports Supported Research Project. The second author would like to thank Alec Jacobson for insightful conversations on task-specific uncertainty quantification. 
AT was supported by Cornell University, jointly via the Center for Data Science for Enterprise and Society, the College of Engineering, and the Ann S. Bowers College of Computing and Information
Science. 
\end{acks}

\bibliographystyle{ACM-Reference-Format}
\bibliography{bibliography}

\balance
\clearpage
\nobalance
\appendix
\renewcommand{\bibliographystyle}[1]{}
\renewcommand{\bibliography}[1]{}

\twocolumn[{
\sffamily\Huge Supplementary Material for "A Bayesian Approach for Task-Specific Next-Best-View Selection with Uncertain Geometry"

\strut
}]

\section{Additional Implementation Details}
\label{sec:detail}

\paragraph{Scanning details:} 
The $\f{scan}$ operation is realized using the differentiable rasterizer in PyTorch3D~\cite{ravi2020accelerating}, which computes ray-mesh intersections for both the ground-truth mesh and the sampled posterior implicit surfaces $\h{f}$.
To facilitate rasterization, we evaluate the sampled implicit functions on a $100^3$ voxel grid and extract surface meshes via marching cubes \cite{lorensen1998marching}.
In all our experiments, we model the scanning sensor as a pinhole camera with a field-of-view (FoV) of $ 30^\circ$. The scanning resolution is $50\times 50$ for the classification and heat diffusion tasks, and $ 64\times 64$ for the segmentation task. The radius of the bounding sphere on which the camera candidates are placed is 2.5.

\paragraph{Stochastic Poisson Surface Reconstruction details:} Throughout all our experiments, we use the geometric Gaussian Process model with a $\nu=\frac{3}{2}$ Matérn kernel on the torus $\mathbb{T}^3$, the same as \citet{holalkere2025stochastic}. The corresponding length scale and amplitude are chosen as 0.005 and 0.2, respectively. Following \citet{holalkere2025stochastic}, posterior sampling in geometric stochastic Poisson surface reconstruction is accelerated by precomputing data-agnostic GP prior samples and applying data-dependent updates at runtime. The precomputed prior samples are stored on a $100^3$ grid within $[-1.75, 1.75]^3$. The truncation threshold for the Karhunen–Loève expansion is $L=50$.
For all proposed methods and baselines that utilize this pipeline, we normalize all scene coordinates to within $[-\frac{\pi}{2}, \frac{\pi}{2}]^3$, as is required. The number of samples used for Monte Carlo estimation of the expected acquisition function is set to $S = 16$.

\begin{table}[b]
\caption{Ablation study between our method with task-specific acquisition and coverage-guided acquisition on the segmentation task.}
\label{tab:recon}
\begin{tabular}{lcc}
\toprule
Method & Chamfer Distance ($\downarrow$) & Hausdorff Distance ($\downarrow$) \\\midrule
Ours & \underline{0.113} & \textbf{0.536} \\
FPS & \textbf{0.109} & \underline{0.573} \\
Uncertainty & 0.145 & 0.642 \\\bottomrule
\end{tabular}
\end{table}

\section{Additional Experiments}

\paragraph{3D reconstruction quality using our coverage-guided acquisition function:} 
We validate that our method is compatible with the traditional 3D reconstruction task, focusing on scene completeness and reducing reconstruction error. 
In this experiment, we use our Chamfer-distance-based acquisition function $\alpha_{f \mid \mathcal{D}_t}^{(\mathrm{CD})}(\theta)$ targeting surface coverage, and compare against FPS and Uncertainty reduction~\cite{holalkere2025stochastic}. 
We subsample 32 scenes from ShapeNet as the testing set. 
We use each algorithm to select 5 views, scan the scene to obtain a point cloud from these views, use the point cloud for Poisson Surface Reconstruction, and finally compare the reconstructed mesh with the ground truth using the Chamfer distance (CD) and Hausdorff distance (HD). 
As shown in \cref{tab:recon}, our approach is competitive with these baselines that solely focus on scene coverage.

\begin{figure}[t]
\includegraphics{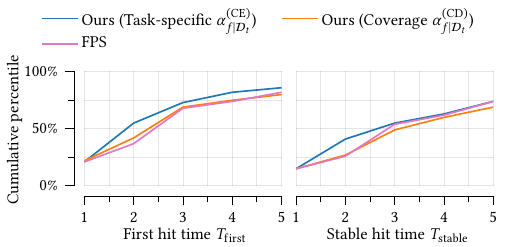}
\caption{Ablation study between our method with task-specific acquisition and coverage-guided acquisition on the classification task of the Truck-ModelNet10 dataset.}
\label{fig:ab_truck}
\end{figure}

\paragraph{Ablation studies between our task-specific acquisition and the general coverage-guided acquisition functions:}
We conduct an ablation study comparing our method using the task-specific acquisition function and the coverage-guided acquisition function $\alpha_{f \mid \mathcal{D}_t}^{(\mathrm{CD})}(\theta)$ to validate the effect of our task-specific design. 
We compare the results in both classification and segmentation tasks, results shown in \cref{fig:ab_truck,tab:ab_seg}. 
From the results, our method with task-specific acquisition significantly outperforms the coverage-guided approach. 
Notably, for the segmentation task, the coverage-guided approach even yields a worse result than random search. 
We see that solely focusing on geometry coverage ignores scene semantics, prioritizing large unseen geometry regions, rather than smaller unseen but semantically important regions.

\begin{table}[b]
\caption{The average number of cameras required for discovering all parts of the object over 80 test shapes, under different $N_{\f{target}}$ criteria. The first two rows are the same as those in Table 1 of our main paper. Less is better.}
\label{tab:ab_seg}
\begin{tabular}{lccccc}
\toprule
$N_{\f{target}}$ & 20 & 40 & 60 & 80 & 100 \\\midrule
Ours (Task-specific) & \textbf{2.93} & \textbf{3.16} & \textbf{3.43} & \textbf{3.67} & \textbf{3.75} \\
Random & 3.29 & 3.42 & 3.65 & 3.93 & 4.10 \\
{Ours (Coverage)} & {3.23} & {3.46} & {3.75} & {4.11} & {4.34} \\
\bottomrule
\end{tabular}
\end{table}  

\begin{figure}[h]
\includegraphics{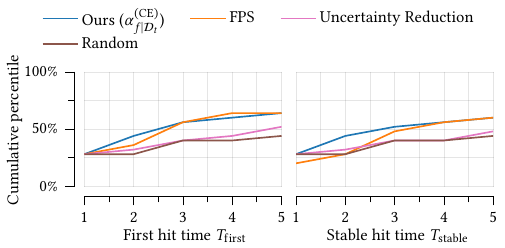}
\caption{Comparison on the classification of the OmniObject3D dataset.}
\label{fig:omni}
\end{figure}

\paragraph{Additional results on the OmniObject3D~\cite{wu2023omniobject3d} dataset:}
We conduct an additional experiment of classification on the OmniObject3D dataset, which is a more diverse, complex, high-fidelity 3D object dataset compared to ModelNet10. Specifically, we select 5 classes \emph{apple}, \emph{lemon}, \emph{mango}, \emph{mushroom}, \emph{orange} within the dataset, train a PointNet++ network as a point cloud classifier, and run our NBV selection algorithm using the cross-entropy-based acquisition function.
We select 5 scenes per class (25 scenes in total) for the test set.
We compare our method against FPS and Uncertainty reduction~\cite{holalkere2025stochastic} on the first hit time and stable hit time metrics.
As shown in \cref{fig:omni}, our method achieves faster convergence.

\bibliographystyle{ACM-Reference-Format}
\bibliography{bibliography}

\end{document}